# Web portals, instant messaging and web communities: new tools for online collaboration

R. Esposito, P. Mastroserio, G. Tortone
*INFN, Napoli, I-80126, Italy*

F. M. Taurino
*INFM, Napoli, I-80126, Italy*

Web portals are nowadays very popular. This type of web sites let users to share information, request advice or help in a particular field, and furthermore allow them to create and extend sites content. Combined with instant messaging systems, used to send messages or files instantaneously to a user or a group of users, and with various kinds of chat programs, which connect two or more individuals simulating a conversation, it is now possible to create "web communities". In this paper we present the opensource tools used to create the "Grid Support Community" of the National Institute for Nuclear Physics in Italy.

## 1. GRIDSUPPORT

The INFNGRID [1] support infrastructure has been implemented in order to offer help and answers to various type of requests and questions. This service is offered to all the members of our community to enable in more sites and experiments the installation, configuration and usage of the GRID.

A web portal is a web site used to collect information on the Net. It allows users to reach resources and documents through a single access point; it would not be easy to find these information elsewhere on Internet. Nowadays our support site is http://gridsupport.na.infn.it.

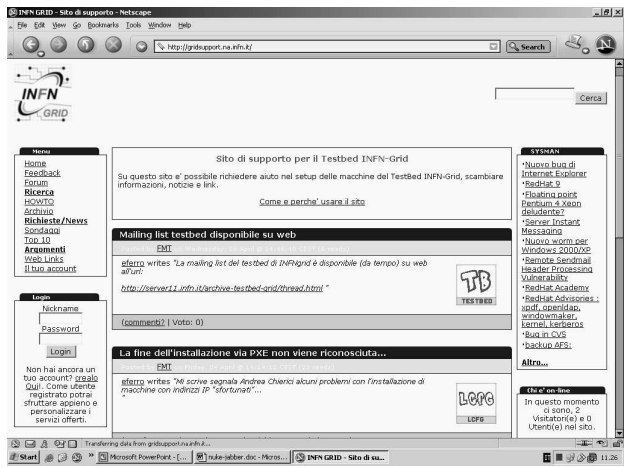

## 2. WEB PORTALS IN MINUTES

### 2.1. PHP-Nuke

We have chosen PHP-Nuke [2], an open source portal system, to implement this site. This package, written in PHP [3], works in combination with the Apache [4] web server and the MySql [5] database, where all documents, requests and answers are stored.

PHP-Nuke allows our users to:
- insert requests, news, answers and comments via form
- search in our knowledge base and documents
- access forums and surveys
- create some content, like mini-howto, *directly online*

The knowledge base actually is organized in about 20 topics, like "LCFG", "Examples", "Software configuration". In this way users can restrict the field of their requests or search only in a specific topic.

Once a request is posted on Gridsupport, other users or members of support team can read the question via a browser and post a reply or a comment, extending the knowledge base. Users can search through documents, requests and answers posted on the site. There are also some specific forums, relative to the "use cases" for examples, and some surveys, used to verify the opinions of our users in fields like languages, ease of installation, most important problems. Important documents are published or linked on the site, and several of them can be composed and modified online via a browser.

## 3. INSTANT MESSAGING

Instant Messaging (IM) enables a user to determine the online availability of another user, allows them to instantly exchange messages and combines the urgency of a phone call with the functionality of e-mail.

Most of the popular instant-messaging systems provide a variety of features:
- Instant messages - Send notes to a user who is online
- Chat - Create your own custom chat room with co-workers or friends
- Files - Share files by sending them directly to your co-workers
- Talk - Use the Internet instead of a phone to actually talk with other users

### 3.1. Jabber

We have installed an IM server using the GPLed server Jabberd [6], which implements the Jabber protocol, based on XML, for the real-time exchange of messages and presence between any two points on the Internet.

We have chosen to implement a private instant messaging system with Jabber because it is open. No company controls the protocol, and anybody can write or has written plug-ins and clients for it. Jabber can also communicate with all the major proprietary IM systems through public gateways. It is also possible to have an IM address similar to an email address (username@jabber.yourdomain.org), rather than something like 19015012 or some strange private IM screen name. Furthermore clients are available for any platform and users' buddy lists and preferences are stored on the server.

## 4.  CONCLUSIONS

The INFNGRID support website, is only an example of a powerful collaboration tool that can be easily implemented combining opensource software, such as PHP-Nuke, Jabber and the always present mailing list. Thanks to these tools any support team can offer a pervasive presence to a "web community" in order to guarantee information exchange and fast response to the problems.